\documentclass[a4paper]{aa}

\usepackage{graphicx,natbib,txfonts}

\newcommand{\Bz}{$\langle B_\mathrm{z} \rangle$}
\newcommand{\kms}{km\,s$^{-1}$}
\newcommand{\Teff}{$T_\mathrm{eff}$}
\newcommand{\vsini}{$v_\mathrm{e}\sin{i}$}
\newcommand{\s}{$\varphi$~Phe}

\begin{document}

   \title{Magnetism, chemical spots, and stratification in \\the HgMn star $\varphi$~Phoenicis\thanks{Based on observations collected at the European Southern Observatory, Chile (ESO programme 084.D-0338).}}

   \subtitle{}

   \author{V.~Makaganiuk\inst{1}
      \and O.~Kochukhov\inst{1}
      \and N.~Piskunov\inst{1}
      \and S.~V.~Jeffers\inst{2}
      \and C.~M.~Johns-Krull\inst{3}
      \and C.~U.~Keller\inst{2}
      \and M.~Rodenhuis\inst{2}
      \and F.~Snik\inst{2}
      \and H.~C.~Stempels\inst{1}
      \and J.~A.~Valenti\inst{4}}

   \institute{Department Physics and Astronomy, Uppsala University, Box 516, 751 20 Uppsala, Sweden
   \and
Sterrekundig Instituut, Universiteit Utrecht, P.O. Box 80000, NL-3508 TA Utrecht, The Netherlands
   \and
Department of Physics and Astronomy, Rice University, 6100 Main Street, Houston, TX 77005, USA
   \and
Space Telescope Science Institute, 3700 San Martin Dr, Baltimore MD 21211, USA}

   \date{Received 28 Sep 2011 / Accepted 25 Nov 2011}

  \abstract
   {Mercury-manganese (HgMn) stars have been considered as non-magnetic and non-variable chemically peculiar (CP) stars for a long time. However, recent discoveries of the variability in spectral line profiles have suggested an inhomogeneous surface distribution of chemical elements in some HgMn stars. From the studies of other CP stars it is known that magnetic field plays a key role in the formation of surface spots. All attempts to find magnetic fields in HgMn stars have yielded negative results.}
   {In this study, we investigate the possible presence of a magnetic field in \s\ (HD~11753) and reconstruct surface distribution of chemical elements that show variability in spectral lines. We also test a hypothesis that a magnetic field is concentrated in chemical spots and look into the possibility that some chemical elements are stratified with depth in the stellar atmosphere.}
   {Our analysis is based on high-quality spectropolarimetric time-series observations, covering a full rotational period of the star. Spectra were obtained with the HARPSpol at the ESO 3.6-m telescope. To increase the sensitivity of the magnetic field search, we employed  the least-squares deconvolution (LSD) technique. Using Doppler imaging code INVERS10, we reconstructed surface chemical distributions by utilising information from multiple spectral lines. The vertical stratification of chemical elements was calculated with the DDAFit program.}
   {Combining information from all suitable spectral lines, we set an upper limit of 4~G on the mean longitudinal magnetic field. For chemical spots, an upper limit on the longitudinal field varies between 8 and 15~G.  We confirmed the variability of Y, Sr, and Ti and detected variability in Cr lines. Stratification analysis showed that Y and Ti are not concentrated in the uppermost atmospheric layers.}
  {Our spectropolarimetric observations rule out the presence of a strong, globally-organised magnetic field in \s. This implies an alternative mechanism of spot formation, which could be related to a non-equilibrium atomic diffusion. However, the typical time scales of the variation in stratification predicted by the recent time-dependent diffusion models exceed significantly the spot evolution time-scale reported for \s.}

   \titlerunning{Magnetism, spots, and stratification in \s}
   \maketitle


\section{Introduction}
\label{intro}

Mercury-manganese (HgMn) stars is one of the sub-classes of early-type chemically peculiar (CP) stars. HgMn stars show an overabundance of Hg, Mn, Y, Sr, and other, mostly heavy, chemical elements. These stars are often found in close binaries and belong to the temperature range of \Teff\,=\,9500--16000~K, corresponding to the spectral classes from A0 to B5 \citep{Dworetsky:1993}.

Earlier it was thought that HgMn stars show no rotational spectral variability, meaning that the atmospheres of these stars are horizontally homogeneous. However, the variability discovered in a spectral line of Hg in $\alpha$~And \citep{Adelman:2002} has changed this picture. \citet{Adelman:2002} show that the observed variability is related to presence of chemical spots. In the case of Ap stars, the presence of spots is usually caused by the magnetic field. The study by \citet{Wade:2006a} aimed at the search for magnetic field in $\alpha$~And proved an absence of the global magnetic field sufficient to cause the chemical spot formation.

Detection of spots increased the interest towards HgMn stars, resulting in a discovery of seven more spotted HgMn stars: HR~1185 and HR~8723 \citep{Kochukhov:2005}, AR~Aur \citep{Hubrig:2006, Folsom:2010}, HD~11753, HD~53244, HD~221507 \citep{Briquet:2010}, and HD~32964 \citep{Makaganiuk:2011b}. At the same time, all systematic studies of the magnetic field in HgMn stars reported its absence \citep{Shorlin:2002, Wade:2006a, Folsom:2010, Auriere:2010, Makaganiuk:2011a}. In the best cases, an upper limit set on the strength of the mean longitudinal magnetic field is a few Gauss \citep{Auriere:2010, Makaganiuk:2011a}.

The source \s\ (HD~11753, HIP~8882, HR~558) is a slowly-rotating, bright (V$\,=\,5.11$) and cool \citep[\Teff$=10700$~K,][]{Smith:1993a} HgMn star. \citet{Dworetsky:1982} suggested that the radial velocity of \s\ varies with a period of more than 30 days, which means that this star is a spectroscopic binary with a single spectrum. \citet{Leone:1999} found the orbital period of $41.489\pm0.019$~d for \s.

Employing the IUE spectra and the spectrum synthesis method, \citet{Smith:1993a, Smith:1993b, Smith:1994, Smith:1997} determined abundances of Cr, Mn, Fe, Mg, Al, Si, Zn, Cu, Co, Ni, and Hg. \citet{Jomaron:1999} and \citet{Dolk:2003} used optical spectra to determine abundances of Mn and Hg, respectively.

\citet{Smith:1996} found a stratification of gallium, based on the analysis of the UV lines of this element. He concluded that Ga shows an increase of concentration, starting from $\log\tau_{5000}\,=\,+0.3$ towards the upper layers of the stellar atmosphere.

\citet{Briquet:2010} discovered chemical spots on \s. Using a large number of observations, the authors found variability in the spectral lines of Ti, Sr, and Y with a 9\fd54 period. They reconstructed surface maps for the two sets of spectra of this star, separated by 65 days. Based on the differences in their maps, the authors suggest an evolution of spots, similar to the phenomenon discovered by \citet{Kochukhov:2007} in $\alpha$~And.

There were no previous magnetic field studies of \s. This star was included in the sample of HgMn stars for which we performed a magnetic field survey with HARPSpol \citep{Makaganiuk:2011a}. The spectropolarimetric observations of \s\ cover its full rotational period, enabling us to measure the magnetic field at each rotational phase. Based on these high-quality data we also reconstructed surface maps of variable chemical elements and tested the presence of extreme vertical stratification for some of them.

In Sect.~\ref{obs} we describe the observations and data reduction. The least-squares deconvolution (LSD) and the measurements of the longitudinal magnetic field are presented in Sect.~\ref{mf}. Section~\ref{lpv} describes investigation of the line profile variability. Surface maps of chemical elements derived with the Doppler imaging technique are discussed in Sect.~\ref{DI}. The stratification analysis is presented in Sect.~\ref{strat}. Section~\ref{disc} summarises our results and discusses them in the context of previous studies of \s\ and recent theoretical diffusion calculations.

\section{Observations and data reduction}
 \label{obs}
The star \s\ was observed in January 2010, using a newly-built HARPS polarimeter \citep{Snik:2011, Piskunov:2011} at the 3.6-m ESO telescope in La Silla, Chile. All observations were done with the circular polarimeter. Data collected during the first two nights were part of the engineering run devoted to the commissioning of the instrument. Orientation of the quarter-wave retarder plates was not calibrated, therefore no useful Stokes $V$ spectra could be obtained. During the ten other nights, we obtained complete spectropolarimetric observations of \s. A typical signal-to-noise ($S/N$) ratio of our spectra is 300\,--\,400 per CCD column at $\lambda\approx5200$~\AA\ and the resolving power is $\lambda/\Delta\lambda=115000$.

Spectra were recorded by a mosaic of two 2K~$\times$~4K CCDs. One CCD records the blue part of a spectrum, which contains 90 spectral orders in the spectropolarimetric mode, while the red part of a spectrum is recorded by the other CCD. The latter contains 52 spectral orders. The overall wavelength coverage is 3780--6920~\AA, with a gap between 5259 and 5337~\AA. For the calibration procedure we used a standard set of frames, which consisted of ten biases, ten flat fields exposed in circular polarisation mode, and one ThAr frame. Such a set was obtained at the beginning and end of each observing night. The polarimetric observations consisted of four sub-exposures, corresponding to 45$\degr$, 135$\degr$, 225$\degr$ or 315$\degr$ position of the quarter-wave plate. The exposure times adopted for individual sub-exposures varied from 270 to 300 seconds, depending on the weather conditions.

The rotational period of \s, 9\fd54 \citep{Briquet:2010}, was recently revised to 9\fd531 \citep{Korhonen:2011}. This improvement corresponds to a negligible change of the rotational phases of our observations. Therefore, we adopted the period published by \citet{Briquet:2010} throughout our analysis. In the Table~\ref{tab1} we list the heliocentric Julian dates of the observations, rotational phase, type of the observations, exposure times, and $S/N$. The rotational phases were computed relative to the first night of observations, assuming the rotational period of 9\fd54 and the zero phase at HJD$\,=\,2455197.06789$.

\begin{table}[t]
   \caption{The log of the HARPSpol observations of \s.}
   \label{tab1}
   \centering
   \begin{tabular}{c c c r r}
  \hline\hline
HJD$-24\times 10^{5}$ & Phase & Stokes & $t_\mathrm{exp}$ (s) & $S/N$ \\
  \hline
55197.06789 & 0.000 & I  & 4$\times$48  & 190 \\
55198.04517 & 0.102 & I  & 4$\times$47  & 183 \\
55198.05107 & 0.103 & I  & 4$\times$200 & 507 \\
55198.06359 & 0.105 & I  & 4$\times$300 & 388 \\
55199.04987 & 0.208 & I  & 4$\times$200 & 352 \\
55199.07179 & 0.210 & I  & 4$\times$41  & 248 \\
55199.09245 & 0.212 & I  & 4$\times$200 & 365 \\
55200.05063 & 0.313 & IV & 4$\times$200 & 299 \\
55201.02209 & 0.415 & IV & 4$\times$200 & 395 \\
55202.03895 & 0.521 & IV & 4$\times$200 & 433 \\
55203.02838 & 0.625 & IV & 4$\times$300 & 348 \\
55204.01590 & 0.728 & IV & 4$\times$280 & 430 \\
55205.04475 & 0.836 & IV & 4$\times$280 & 353 \\
55206.02121 & 0.939 & IV & 4$\times$270 & 406 \\
55207.01430 & 0.043 & IV & 4$\times$270 & 476 \\
55209.00910 & 0.252 & IV & 4$\times$270 & 379 \\
55210.03112 & 0.359 & IV & 4$\times$270 & 564 \\
  \hline
   \end{tabular}
\end{table}

To process our data we employed the REDUCE package \citep{Reduce}, which performs the standard reduction steps, such as averaging of bias and flat field frames and subtracting the master bias from the master flat field and science frames. Each science frame was corrected for the pixel-to-pixel sensitivity variations using a normalised flat field. All stellar spectra were corrected for the scattered light before the optimal extraction.

The wavelength calibration was performed based on one ThAr spectrum for each night, yielding an accuracy of 18--21~m~s$^{-1}$ of the derived wavelengths. Using a single comparison spectrum per night is sufficient for our study, thanks to the intrinsic stability of the HARPS spectrometer \citep{HARPS}. The normalisation of spectra was done first by correcting the blaze shape and then dividing by a smooth function that was fitted to an upper envelope of the merged spectrum.

To minimise the possible effects of spurious polarisation, we employed the ratio method \citep{Bagnulo:2009} to combine polarimetric sub-exposures and derive continuum-normalised circular polarisation ($V/I_\mathrm{c}$) spectra. For brevity, we refer to these data as ``Stokes V spectra''. In addition to the polarisation spectra, the ratio method produces a diagnostic spectrum, which is called the null ($N$) spectrum. To derive the null spectrum, the four sub-exposures are combined in a such way that the stellar polarisation is destroyed, enabling us to diagnose possible spurious polarisation.

\section{Magnetic analysis}
 \label{mf}
 \subsection{LSD technique}
  \label{LSD}

A search for weak magnetic fields requires polarimetric data of a very high $S/N$. Achieving this directly is technically challenging and costly in terms of the observational time. On the other hand, the circular polarisation profiles are very similar in shape for the majority of spectral lines, especially in weak fields. Therefore, an alternative approach to increasing polarimetric sensitivity while maintaining reasonable exposure times is to use a multi-line diagnostic technique, such as the least-squares deconvolution (LSD) \citep{Donati:1997,Kochukhov:2010a}.

This technique assumes that all spectral lines are identical in shape and can thus be described by a scaled mean profile. If we assign a certain set of weights to individual spectral lines, then the mean profile multiplied by these weights should reproduce an observed spectrum. Utilising observations and a set of weights, LSD code computes a mean profile, combining information from many spectral lines. As a result, the $S/N$ increases by a factor of up to 30\,--\,40, depending on the number of lines in a stellar spectrum. LSD has proven to be a powerful tool for detecting weak magnetic fields \citep{Auriere:2009, Lignieres:2009}. For HgMn stars, it allowed for an upper limit to be established for the mean longitudinal magnetic field at the level of $\le10$~G \citep{Auriere:2010, Makaganiuk:2011a} in the best cases.

\subsection{LSD mask and stellar parameters}
  \label{sp&mask}

LSD technique requires a list of spectral line positions, their central intensities, and magnetic sensitivity, collectively called a line mask. To compile such a mask, it is important to adopt correct stellar parameters since the intensities and the total number of spectral lines extracted from the Vienna Atomic Line Database \citep[VALD,][]{VALD,Kupka:1999} will depend on the effective temperature and chemical composition of a star.

We estimated \Teff\ and $\log g$ from the Str\"omgren photometry \citep{Hauck:1998} with the help of the TempLogG code \citep{Kaiser:2006}, using the calibration by \citet{Moon:1985}. The resulting \Teff\,=\,10500~K and $\log g\,=\,3.8$ agree with the parameters determined by \citet{Smith:1993a} (\Teff$\,=\,10700$~K, $\log g\,=\,3.8$) and \citet{Dolk:2003} (\Teff$\,=\,10612$~K, $\log g\,=\,3.79$). The scatter of the effective temperature determinations suggests \Teff\ uncertainty around 200~K. According to \citet{Moon:1985}, the uncertainty of the effective temperature and surface gravity established using their calibration is $\pm$260~K and $\pm$0.1~dex, respectively. The model atmosphere of \s\ was computed with the LLmodels code \citep{LLmodels} for \Teff$\,=\,10500$~K and $\log g\,=\,3.8$.

The abundances of many chemical elements in \s\ are unknown, although some other elements have been studied previously (see Sect.~\ref{intro}). Using these individual abundance and adopting average values typical of HgMn stars for elements not analysed before, we compiled a preliminary line list and computed synthetic spectrum with the SYNTH3 code \citep{SYNTH3}. Chemical abundances were refined by comparing this calculation to the observed spectrum obtained at HJD$\,=\,$55210.03112, which has the highest $S/N$. Zero microturbulent velocity was assumed both for the VALD extraction and for the spectrum synthesis.

We found good agreement between our observations and the synthetic spectrum for the spectral lines of chemical elements studied previously. However, since we adopted a somewhat cooler model atmosphere compared to \citet{Smith:1993a}, we had to reduce the abundances of Fe, Cr, Mg, Mn, and Ga by up to $\approx$\,0.4~dex. No changes were required for Si and Co. Relatively large corrections were introduced for Sc, Y, Pt, and Zr, because spectral lines of some of these chemical elements were excessively strong or too weak, mismatching the observations. Since the precise quantitative abundance analysis is beyond the scope of our study, we adjusted abundances of individual chemical elements visually, without deriving formal error bars. This is sufficient for compiling an LSD line mask. The abundances are given in the Table~\ref{tab2} as follows: chemical element and the initial and final abundances. The revised abundances were used to obtain a new line list from VALD, which provided 3827 spectral lines in the 3781--6915~\AA\ range.

\begin{table}[t]
   \caption{Abundances of chemical elements in \s.}
   \label{tab2}
   \centering
   \begin{tabular}{c c c}
  \hline\hline
Element & Initial abundance & Final abundance\\
  \hline
He & $-2.07$\tablefootmark{\dag} & $-1.65$ \\
N  & $-4.26$\tablefootmark{\dag} & $-4.46$ \\
O  & $-3.58$\tablefootmark{\dag} & $-3.45$ \\
Na & $-5.87$\tablefootmark{\dag} & $-5.35$ \\
Mg & $-4.64$\tablefootmark{2}    & $-5.00$ \\
Al & $-7.20$\tablefootmark{2}    & $-7.20$ \\
Si & $-4.68$\tablefootmark{2}    & $-4.70$ \\
P  & $-5.68$\tablefootmark{\dag} & $-6.00$ \\
S  & $-5.40$\tablefootmark{\dag} & $-5.20$ \\
Ca & $-5.53$\tablefootmark{\dag} & $-5.73$ \\
Sc & $-8.49$\tablefootmark{\dag} & $-10.75$\\
Ti & $-6.64$\tablefootmark{\dag} & $-6.50$ \\
V  & $-8.04$\tablefootmark{\dag} & $-9.54$ \\
Cr & $-5.80$\tablefootmark{1}    & $-6.00$ \\
Mn & $-5.80$\tablefootmark{5}    & $-5.90$ \\
Fe & $-4.35$\tablefootmark{1}    & $-4.45$ \\
Co & $-9.50$\tablefootmark{1}    & $-9.50$ \\
Ni & $-6.90$\tablefootmark{1}    & $-6.90$ \\
Cu & $-7.32$\tablefootmark{3}    & $-7.32$ \\
Zn & $-10.00$\tablefootmark{3}   & $-10.00$ \\
Ga & $-7.10$\tablefootmark{4}    & $-7.30$ \\
Sr & $-7.12$\tablefootmark{\dag} & $-7.35$ \\
Y  & $-7.83$\tablefootmark{\dag} & $-6.95$ \\
Zr & $-7.45$\tablefootmark{\dag} & $-8.50$ \\
Pt & $-7.66$\tablefootmark{\dag} & $-6.75$ \\
Hg & $-6.88$\tablefootmark{6}    & $-6.88$ \\
  \hline
   \end{tabular}
   \tablefoot{References to the initial abundance are marked as follows:
   \tablefoottext{\dag}{average abundance estimate for HgMn star,}
   \tablefoottext{1}{\citet{Smith:1993a},}
   \tablefoottext{2}{\citet{Smith:1993b},}
   \tablefoottext{3}{\citet{Smith:1994},}
   \tablefoottext{4}{\citet{Smith:1996},}
   \tablefoottext{5}{\citet{Smith:1997},}
   \tablefoottext{6}{\citet{Jomaron:1999}.}}
\end{table}

To apply LSD, we constructed a line mask containing central wavelengths of spectral lines and a set of weights for Stokes~$I$ and $V$. Central depths returned by VALD were used as weights for the intensity spectrum. For the circular polarisation spectrum, the weights were calculated as $w_V\,=\,d\lambda \overline{g}/\lambda_0$, where $d$ is the central depth of a spectral line, $\lambda$ the wavelength, and $\overline{g}$ the corresponding effective Land\'e factor. The normalisation parameter $\lambda_0=4696$~\AA\ was chosen to be close to the averaged wavelength of all spectral lines in the  mask.

To get optimal results from the LSD technique, it is essential not to include very weak spectral lines, because an extra noise might be added to the LSD profiles. To exclude weak lines, we apply a cutoff criterion, which sets a threshold for the line intensity. All spectral lines weaker than the cutoff level are excluded from the LSD mask. For \s\ we set the cutoff criterion to 0.05, which means that our LSD code \citep{Kochukhov:2010a} will use only those lines that are deeper than 5\% relative to the continuum level in the unbroadened spectrum. This yielded 962 spectral lines for computing LSD profiles.

\subsection{Analysis of LSD Stokes $V$ profiles}
  \label{Bz}
\begin{figure}[!t]
  \centering
 {\resizebox{\hsize}{!}{\rotatebox{90}{\includegraphics{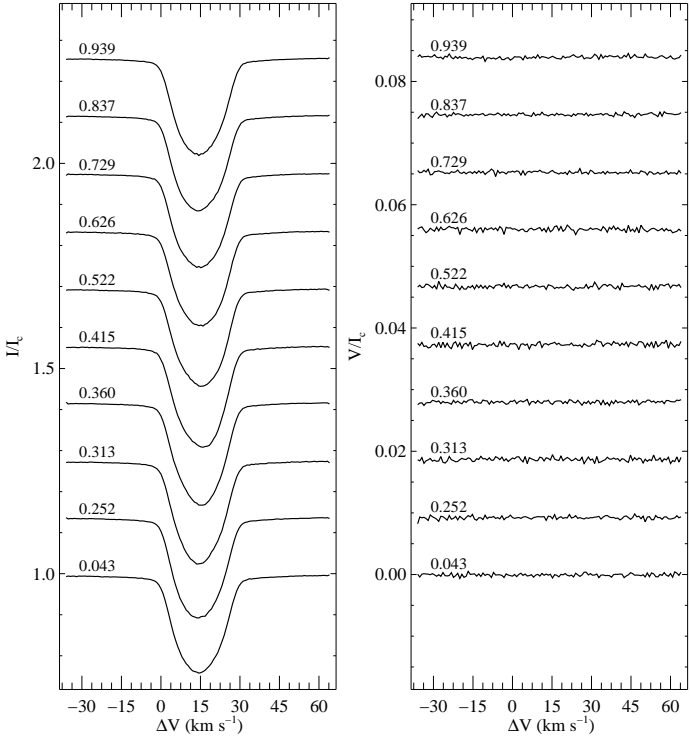}}}}
  \caption{LSD profiles of \s\ for Stokes~$I$ (left panel) and Stokes~$V$ (right panel) spectra. For display purposes we shifted profiles vertically. The rotational phase for each profile is denoted on the left.}
  \label{LSDIV}
\end{figure}

Taking $V_\mathrm{rad}\,=\,14$~\kms\ of \s\ into account, we reconstructed LSD profiles for the velocity range between $-36$ and 64~\kms, which makes the profile extension symmetric with respect to the line centre. An average pixel of the HARPS CCDs has a resolution of 0.8~\kms. We adopted this value as a step for LSD profiles. The $S/N$ reported for the final LSD profiles correspond to this velocity bin. The average $S/N$ gain is nine. The LSD Stokes~$I$ and $V$ profiles for ten rotational phases are shown in Fig.~\ref{LSDIV}. Based on visual examination of the LSD~$V$ profiles, we conclude that no polarisation signal is present.

We estimated the mean longitudinal magnetic field,\Bz, from the first moment of the LSD Stokes~$V$ profile \citep{Kochukhov:2010a}, setting the mean Land\'e factor and central depth to unity, in accordance with the adopted normalisation of LSD weights. All profiles were integrated from $-4$ to 34~\kms. These limits were chosen symmetrically with respect to the centre of the LSD~$I$ profile and were utilised for \Bz\ measurements in the LSD~$V$ and LSD~$N$ profiles. The use of the null profile is an important diagnostics of \Bz\ measurements in LSD~$V$. Should a spurious polarisation alter Stokes~$V$ spectrum, it will be indicated by the \Bz(N) measurements significantly different from zero.

Our analysis of the \Bz(N) measurements showed the absence of a spurious polarisation signal. At the same time, none of \Bz(V) measurements indicates there is a magnetic field in \s. The average error of \Bz\ is 3.8~G and the minimum is 2.8~G.

In addition to the longitudinal field measurements, we performed a false alarm probability (FAP) analysis. This method, based on $\chi^{2}$ statistics, diagnoses a statistical significance of the deviation of the LSD profile from zero. The FAP analysis is especially sensitive to complex magnetic field configurations, which may yield zero longitudinal field. We interpreted FAP results following \citet{Donati:1997}. Any FAP larger than $10^{-3}$ is considered as no detection of the magnetic field. Our FAP values are always greater than 0.1, from which we conclude that \s\ significant tangled magnetic fields.

We summarise the longitudinal magnetic field and FAP measurements in Table~\ref{tab3}. The first column gives the rotational phase, the second column gives $S/N$ in the LSD Stokes~$V$ profiles, while the last four provide longitudinal magnetic field measurements for LSD~$V$ and $N$ with the corresponding FAP numbers.

\begin{table}[t]
   \caption{Magnetic field measurements of \s.}
   \label{tab3}
   \centering
   \begin{tabular}{c c r c| r c}
  \hline\hline
      &            & \multicolumn{2}{c}{LSD~$V$} & \multicolumn{2}{c}{LSD~$N$} \\
Phase & $S/N$(LSD) & \Bz\ (G) & FAP              & \Bz\ (G) & FAP              \\

  \hline
0.043 & 2532 & $-9.7\pm5.1$ & 1.00 & $-9.0\pm5.1$ & 1.00 \\
0.252 & 4069 & $-3.5\pm3.2$ & 0.66 & $-5.4\pm3.2$ & 0.89 \\
0.313 & 3130 & $6.6\pm4.0$  & 0.25 & $4.4\pm4.0$  & 0.49 \\
0.359 & 4357 & $1.3\pm2.9$  & 0.60 & $3.7\pm2.9$  & 0.00 \\
0.415 & 4417 & $1.9\pm2.8$  & 0.16 & $-3.7\pm2.8$ & 0.23 \\
0.521 & 3053 & $-6.0\pm4.1$ & 0.78 & $-1.4\pm4.1$ & 0.68 \\
0.625 & 3404 & $6.8\pm3.8$  & 0.80 & $-1.6\pm3.8$ & 0.47 \\
0.728 & 3311 & $-1.3\pm4.0$ & 0.36 & $1.8\pm3.9$  & 0.66 \\
0.836 & 3954 & $3.3\pm3.3$  & 0.78 & $2.9\pm3.3$  & 0.32 \\
0.939 & 4482 & $-0.7\pm2.9$ & 0.97 & $-3.8\pm2.9$ & 0.26 \\
  \hline
   \end{tabular}
\end{table}

\begin{figure}[!t]
  \centering
 {\resizebox{\hsize}{!}{\rotatebox{90}{\includegraphics{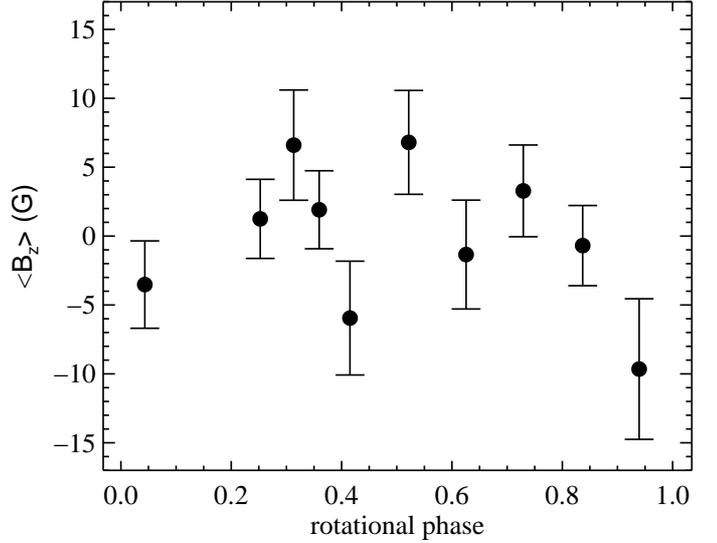}}}}
  \caption{Measurements of the longitudinal magnetic field as a function of rotational phase.}
  \label{Bz_phase}
\end{figure}

To check for a systematic rotational modulation of \Bz, we plotted our measurements as a function of the rotational phase (Fig.~\ref{Bz_phase}). We conclude that there is no evidence of periodicity. All measurements are consistent with zero longitudinal magnetic field.

\begin{figure}[!t]
  \centering
 {\resizebox{\hsize}{!}{\rotatebox{90}{\includegraphics{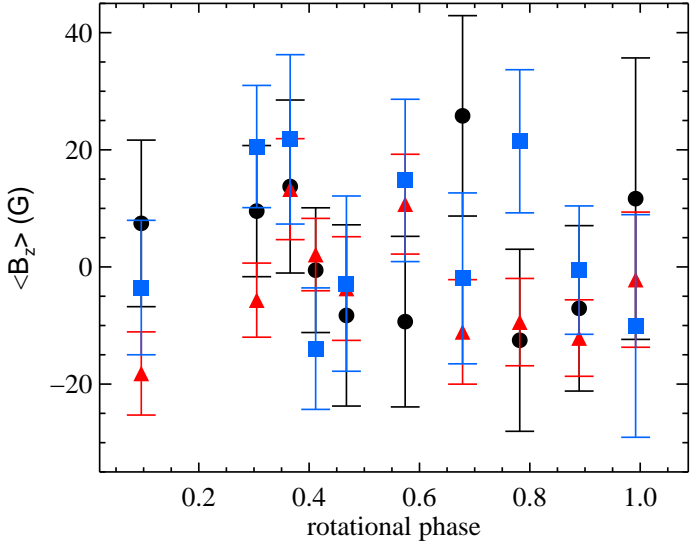}}}}
  \caption{Measurements of the longitudinal magnetic field as a function of rotational phase, based on spectral lines of Y (circles), Ti (triangles), and Cr (squares).}
  \label{Bz_phase_lines}
\end{figure}

It has been suggested by \citet{Hubrig:2010} that magnetic fields in HgMn stars may be concentrated in spots of chemical elements, implying that the polarisation signal should be stronger in the variable spectral lines. To constrain the longitudinal magnetic field in chemical spots, we computed mean profiles for Y, Ti, and Cr using the multi-profile version of our LSD code \citep[see Sect. 2.4.1 in][]{Kochukhov:2010a}. From these LSD profiles, we measured \Bz, finding no magnetic field with a typical error bar of 8~G for Ti, 13~G for Cr, and 15~G for Y lines. These measurements are plotted as a function of rotational phase in Fig.~\ref{Bz_phase_lines}.

Summarising magnetic field analysis results for the inhomogeneously distributed chemical elements, we conclude that there is no evidence of any polarisation signal in their lines.

\section{Line profile variability}
 \label{lpv}
For the line profile variability search, we utilised all available intensity spectra from our observing run. We examined the spectral range from from 3830 to 6830~\AA, ignoring all stellar spectral lines blended by telluric absorption. In total, we found 83 variable spectral lines.

The strongest variability is seen in the lines of Y and Sr. Distortion of the shape of these lines is clearly evident in individual spectra. A somewhat weaker variability is observed for Ti. We see a different variability in spectral lines of various chemical elements, but a similar pattern for the lines of the same element. This type of variability is typical of the spotted distribution of chemical elements. No definite variability was found in individual spectral lines belonging to elements other than Y, Sr, and Ti. Compared to other CP stars, the HgMn type often shows a rather weak variability, which is barely detected based on the visual examination of individual spectral lines. Thus, a numerical assessment and multi-line analysis are necessary for investigating this weak variability.

\begin{figure*}[!t]
  \centering
 {\resizebox{15cm}{!}{\rotatebox{90}{\includegraphics{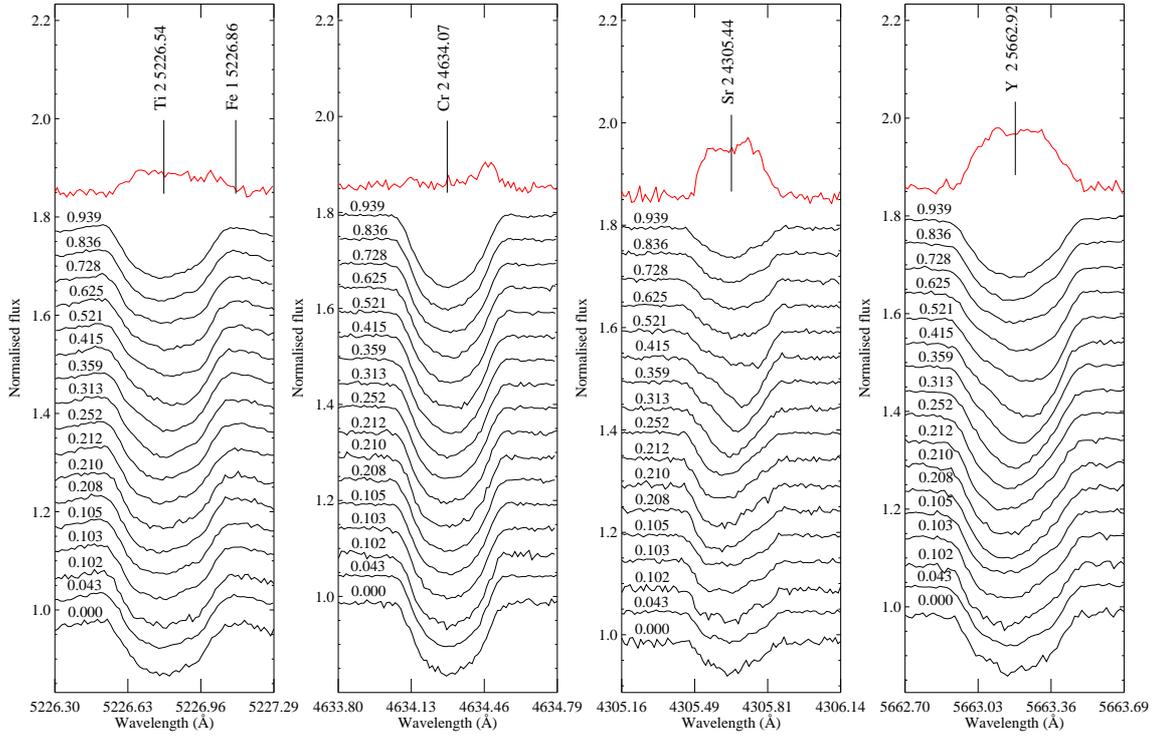}}}}
  \caption{The variability of spectral line profiles in \s. The uppermost curve is a standard deviation computed for all spectra below. For displaying purposes, spectra are shifted vertically.}
  \label{spVar}
\end{figure*}

We computed a standard deviation for all spectral lines. Examination of these data confirmed the variability in Ti, Sr, and Y, but also indicated a marginal variability in Cr. We show 17 phases for selected spectral lines of Ti, Cr, Sr, and Y in Fig.~\ref{spVar}. The uppermost curve represents the standard deviation, which shows a distinctive profile at the position of variable spectral lines. The amplitude of the standard deviation profiles characterises the strength of variability.

\begin{figure*}[!t]
  \centering
 {\resizebox{13cm}{!}{\rotatebox{90}{\includegraphics{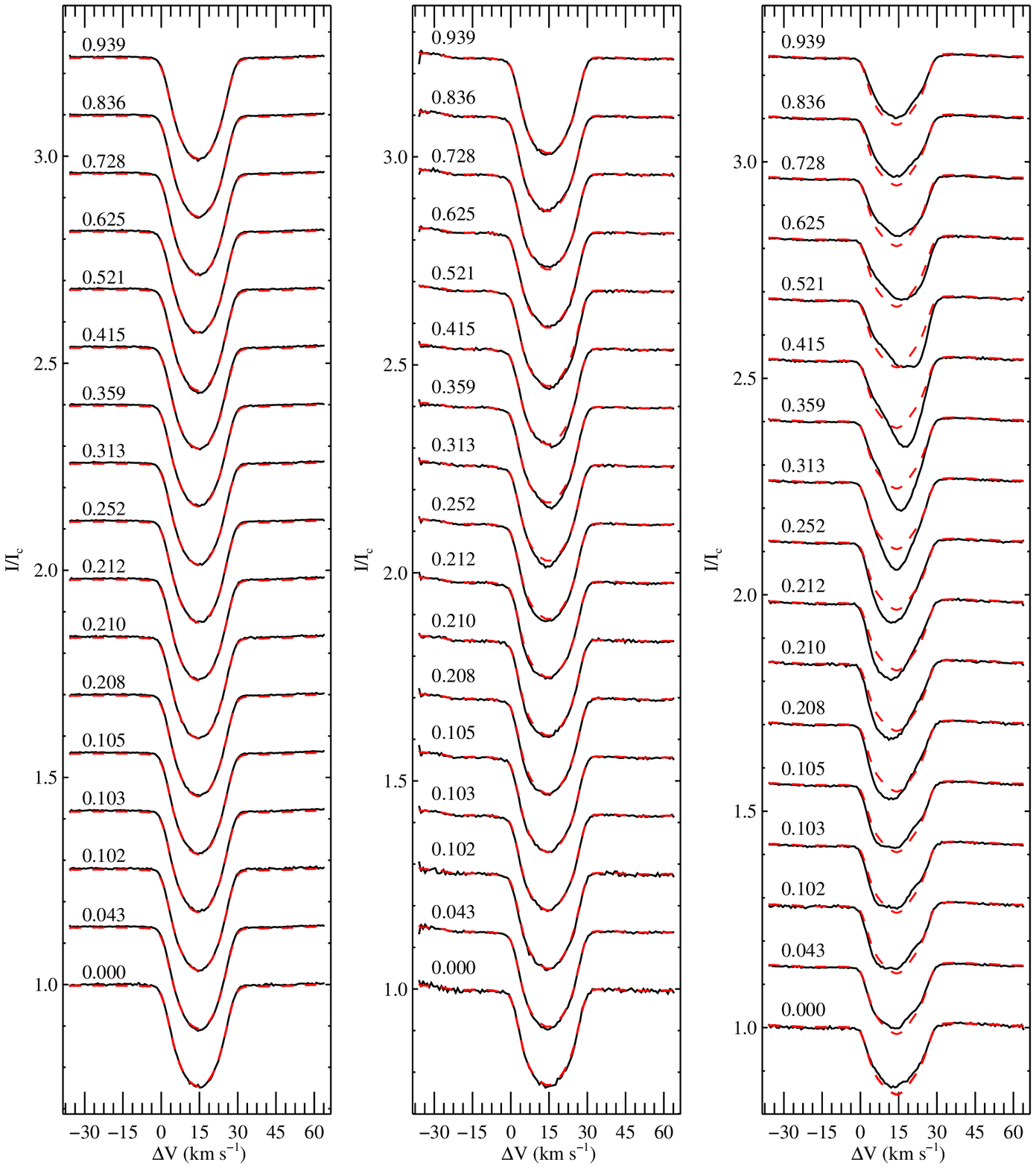}}}}
  \caption{Constant LSD profiles of Fe (left panel), extremely weakly variable Cr profiles (middle panel), and LSD profiles of Y (right panel) as an example of the strongest variability. Dashed line indicates the mean profile. For displaying purposes, profiles corresponding to different rotation phases are shifted vertically.}
  \label{Fig5}
\end{figure*}

We applied the LSD technique to the spectral lines of Cr to verify their variability. Results of this analysis are summarised in Fig.~\ref{Fig5}. For the reference, the LSD profiles for the non-variable chemical element, Fe, are shown in the left-hand panel. The other two panels present LSD profiles for Cr and Y. We computed an average LSD profile for each chemical element and plotted it with dashed lines to highlight the relative profile changes. Indeed, the very weak variability suggested by the standard deviation is seen in the LSD profiles of Cr. The comparison of Fe and Cr profiles allows us to conclude that these weak variations are not introduced by possible instrumental or data-reduction artefacts.

\begin{figure}[!t]
  \centering
 {\resizebox{\hsize}{!}{\rotatebox{90}{\includegraphics{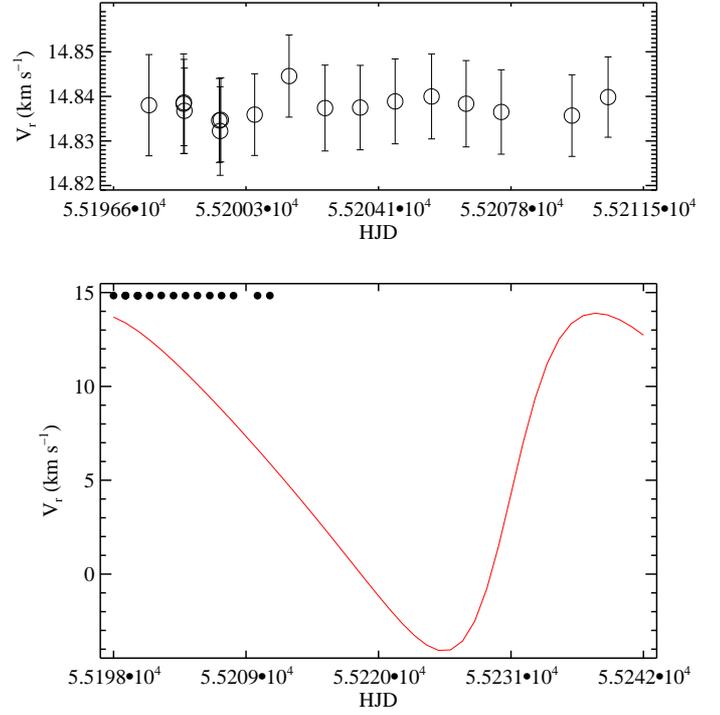}}}}
  \caption{Upper panel: measured radial velocity with respective error bars. Lower panel: the orbital solution (solid line) based on the orbital elements from \citet{Leone:1999} and our measurements (circles) as a function of reduced Julian date.}
  \label{Fig6}
\end{figure}

We mentioned in Sect.~\ref{intro} that \citet{Dworetsky:1982} and \citet{Leone:1999} have suggested that the radial velocity of \s\ changes due to the orbital motion. To verify their results, we used the centre-of-gravity of the LSD~$I$ profiles of Fe to determine the radial velocities for each observation. The resulting mean value of the radial velocity of \s\ is 14.58$\pm$0.02\kms. Individual measurements of radial velocity, together with the respective error bars, are shown in the upper panel of Fig.~\ref{Fig6}. The radial velocity variation predicted by the orbital solution by \citet{Leone:1999} is compared with our measurements in the lower panel of Fig.~\ref{Fig6}. The errors of individual measurements are close to the symbol size.

We conclude that the radial velocity of the star does not change with the suggested 41\fd489 period. If this period were true, we would see a change in radial velocity $\approx$10~\kms\ during our observing run. Moreover, earlier observations of \s\ by \citet{Briquet:2010} with a better time coverage also indicate no change in the radial velocity of this star, yielding the mean radial velocity, 14.24$\pm$0.01~\kms, nearly identical to ours.

\begin{figure}[!t]
  \centering
 {\resizebox{\hsize}{!}{\rotatebox{90}{\includegraphics{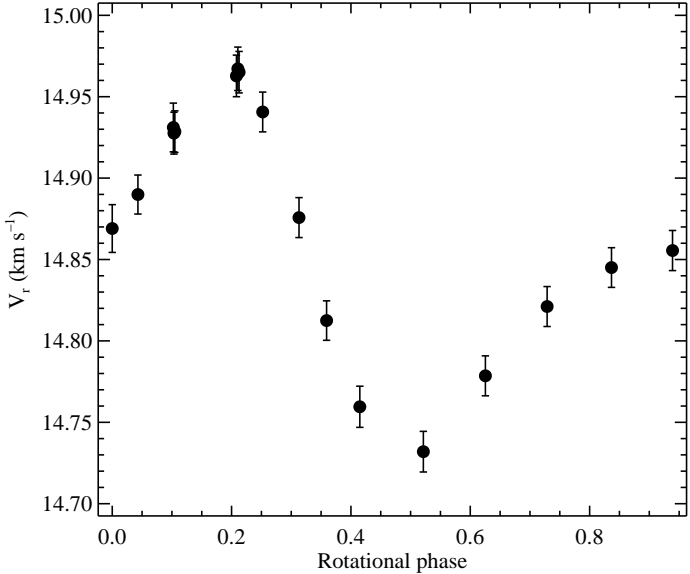}}}}
  \caption{Radial velocities determined from the LSD~$I$ profiles of Y as a function of rotational phase.}
  \label{Fig7}
\end{figure}

We use Y lines, which show the strongest variability compared to other chemical elements, to verify the rotational period of 9\fd54 suggested for \s\ by \citet{Briquet:2010}. The radial velocities derived from the LSD~$I$ profiles of Y (see Fig.~\ref{Fig5}) are shown as a function of rotational phase in Fig.~\ref{Fig7}. Evidently, our data agrees with the previously determined rotational period. A short time span of our observations does not allow deriving any useful independent period estimate.

\section{Doppler imaging}
 \label{DI}
A quantitative interpretation of the observed variations in spectral lines requires using the Doppler imaging (DI) technique, as well as knowledge of physical parameters of a star. One of the key parameters is the projected rotational velocity. To determine \vsini, we selected 20 sufficiently strong unblended spectral lines of iron in the spectrum with the highest $S/N$ (see Table~\ref{tab1}). We fitted the model spectrum with corrected abundances to these observations, using BinMag\footnote{http://www.astro.uu.se/$\sim$oleg/soft.html}. BinMag is an IDL-based widget tool, which allows visual comparison of synthetic spectra with observations, taking radial velocity, instrumental resolution, and rotational and macroturbulent broadening into account. To determine the rotational velocity of \s, the code was allowed to vary \vsini\ for each selected Fe line. The resulting mean value is \vsini$\,=\,$13.62~\kms\ with a standard deviation of 0.22~\kms. We have verified that variation by one standard deviation yields spectrum changes above the noise level, therefore this error estimate is meaningful.

Another parameter required by DI is the inclination angle $i$ between the stellar rotation axis and the line of sight of an observer. It can be determined as
\begin{equation}
  \label{incl}
  \sin i=\frac{P\,v_\mathrm{e}\sin i}{50.613 R},
\end{equation}
where $P$ is the stellar rotational period in days, \vsini\ the projected rotational velocity, and $R$ the radius of a star in solar units. The last is given by the following expression:
\begin{equation}
  \label{rad}
  \frac{R}{R^\odot}=\left(\frac{L}{L^\odot}\right)^{1/2}\left(\frac{T_\mathrm{eff}}{T_\mathrm{eff}^\odot}\right)^{-2}.
\end{equation}
The stellar luminosity can be inferred from the absolute magnitude:
\begin{equation}
  \label{lum}
  \log\left(\frac{L}{L^\odot}\right)=-\frac{M_V+BC-M_{bol}^{\odot}}{2.5}.
\end{equation}
Here $M_V$ is an absolute visual magnitude of the star, $BC$ is the bolometric correction, and $M_{bol}^{\sun}=\,$4.73. The bolometric correction for \s\ was estimated following \citet{Flower:1996}. The absolute visual magnitude was determined as $M_V=V+5+\log{\pi}$. The trigonometric parallax, $\pi\,=\,$10.63$\pm$0.37~mas, was adopted from \citet{van-Leeuwen:2007}. The effective temperature is taken from Sect.~\ref{sp&mask}. From all these parameters we derive the inclination angle $i=65.7\degr\pm7.1\degr$. All parameters of \s\ and their respective errors are summarised in Table~\ref{stPar}.

\begin{table}[t]
   \caption{Physical properties of \s.}
   \label{stPar}
   \centering
   \begin{tabular}{c c}
  \hline\hline
Parameter & Value\\
  \hline
$i$               & $65.7\degr\pm7.1\degr$\\
$M_{v}$           & $0.243\pm0.076$     	\\
$R/R_{\sun}$      & $2.817\pm0.157$     	\\
$\log L/L_{\sun}$ & $1.938\pm0.035$     	\\
$\log g$          & $3.8\pm0.1$			\\
\Teff~(K)         & $10500\pm200$		    \\
\vsini~(\kms)     & $13.62\pm0.22$		\\
  \hline
   \end{tabular}
\end{table}

To interpret the variability of line profiles, we selected 29 spectral lines of Y, Sr, Ti, and Cr. Surface maps were reconstructed using the Doppler imaging code INVERS10 \citep{Piskunov:2002}. Unlike the previous attempt to study chemical spots in \s\ \citep{Briquet:2010}, our analysis is based on simultaneous modelling of multiple spectral lines, which significantly improves reliability of the DI maps. The maps of Y were reconstructed from nine spectral lines, Sr from three, Ti from nine, and Cr from eight spectral lines.

All inversions were constrained using the Tikhonov regularisation \citep[see][]{Piskunov:2002}. The value of the regularisation parameter was chosen in the usual way, by adopting the highest regularisation which provided a fit to the observations compatible with $S/N$ of the data.

Analysing the fit of individual profiles, we found synthetic spectrum deviating systematically for a couple of lines of each element, while the rest of lines were fitted rather well. To minimise the influence of the errors in atomic data, we allowed the code to correct the oscillator strengths of these lines in the course of inversion. In Table~\ref{DIlines} we provide the information about ion, wavelength, excitation potential, initial oscillator strength, and corrected oscillator strength of the spectral lines used in DI inversion.

\begin{table}[t]
   \caption{Spectral lines used for the Doppler imaging of \s.}
   \label{DIlines}
   \centering
   \begin{tabular}{c c c r r}
  \hline\hline
Ion & Wavelength (\AA) & Excit (eV) & Initial $\log gf$ & $\Delta\,\log gf$\\
  \hline
\ion{Y}{ii}  & 4204.695 & 0.000 & $-1.467$ & $-0.226$ \\
\ion{Y}{ii}  & 4398.013 & 0.130 & $-0.895$ &          \\
\ion{Y}{ii}  & 4465.268 & 4.133 &   0.222  & $-0.195$ \\
\ion{Y}{ii}  & 4900.120 & 1.033 &   0.103  &          \\
\ion{Y}{ii}  & 5087.416 & 1.084 & $-0.077$ &          \\
\ion{Y}{ii}  & 5119.112 & 0.992 & $-1.249$ &          \\
\ion{Y}{ii}  & 5497.408 & 1.748 & $-0.276$ &          \\
\ion{Y}{ii}  & 5509.895 & 0.992 & $-0.948$ &   0.076  \\
\ion{Y}{ii}  & 5662.925 & 1.944 &   0.384  &   0.167  \\
\ion{Sr}{ii} & 4077.709 & 0.000 &   0.160  &          \\
\ion{Sr}{ii} & 4215.519 & 0.000 & $-0.140$ &   0.057  \\
\ion{Sr}{ii} & 4305.443 & 3.040 & $-0.130$ & $-0.355$ \\
\ion{Ti}{ii} & 4163.644 & 2.590 & $-0.130$ &          \\
\ion{Ti}{ii} & 4386.844 & 2.598 & $-0.960$ &          \\
\ion{Ti}{ii} & 4395.839 & 1.243 & $-1.930$ &          \\
\ion{Ti}{ii} & 4399.765 & 1.237 & $-1.190$ &          \\
\ion{Ti}{ii} & 4501.270 & 1.116 & $-0.770$ &   0.203  \\
\ion{Ti}{ii} & 4563.757 & 1.221 & $-0.690$ &          \\
\ion{Ti}{ii} & 4779.985 & 2.048 & $-1.260$ &          \\
\ion{Ti}{ii} & 4911.193 & 3.124 & $-0.610$ &          \\
\ion{Ti}{ii} & 5129.152 & 1.892 & $-1.240$ & $-0.035$ \\
\ion{Cr}{ii} & 4037.972 & 6.487 & $-0.670$ & $-0.238$ \\
\ion{Cr}{ii} & 4051.930 & 3.104 & $-2.330$ & $-0.155$ \\
\ion{Cr}{ii} & 4145.781 & 5.319 & $-1.100$ & $-0.114$ \\
\ion{Cr}{ii} & 4275.567 & 3.858 & $-1.730$ &          \\
\ion{Cr}{ii} & 4554.988 & 4.071 & $-1.490$ &   0.052  \\
\ion{Cr}{ii} & 4565.739 & 4.042 & $-1.980$ &          \\
\ion{Cr}{ii} & 4588.199 & 4.071 & $-0.840$ &   0.198  \\
\ion{Cr}{ii} & 4616.629 & 4.072 & $-1.570$ &          \\
  \hline
   \end{tabular}
\end{table}

\begin{figure*}[!t]
  \centering
 {\resizebox{\hsize}{!}{{\includegraphics{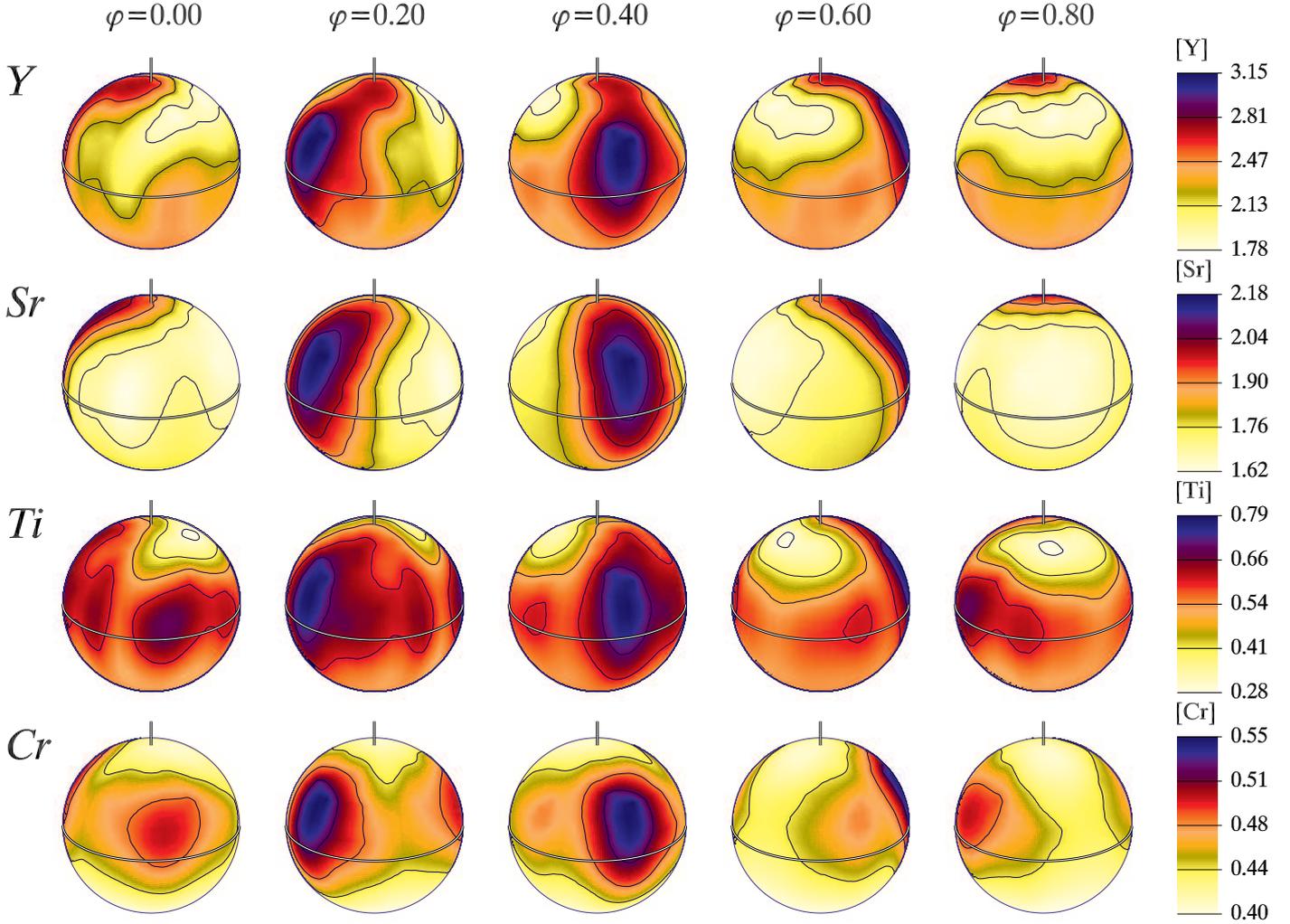}}}}
  \caption{Surface maps of Y, Sr, Ti, and Cr derived for \s. The star is shown at four equidistant rotational phases with rotational axis oriented vertically. Inclination of the rotational axis is 65$\degr$. The abundance scale relative to the Sun is plotted on the right. Darker regions correspond to a higher relative abundance. The contours for Y plotted with a step of 0.33 dex, 0.13 dex for Sr and Ti, and 0.04 dex for Cr.}
  \label{Fig8}
\end{figure*}

We present spherical maps of the distribution of chemical elements in Fig.~\ref{Fig8}. The fit of the observed profiles of Y, Sr, Ti, and Cr is shown in Figs.~\ref{fitY}, \ref{fitSr}, \ref{fitTi}, and \ref{fitCr}, respectively. For \ion{Sr}{ii} line at 4077.71~\AA, we found a much stronger contribution of the blending spectral line of \ion{Cr}{ii} at 4077.51~\AA\ in the synthetic spectrum than appears in observations. The correction of the oscillator strength for this particular Cr line by one order of magnitude allowed a better fit of the observed profile, reducing the standard deviation for Sr from 1.06\% to 0.65\%. For the other chemical elements, the standard deviation of the final profile fit is 0.52\% for Y, 0.45\% for Ti, and 0.51\% for Cr. These results are consistent with the quality of our data.

From the DI surface maps we found that all chemical elements concentrate in the same spot, observed at phases 0.2--0.4. A secondary chemical spot exists for Ti and Cr and is visible at phase range 0.8--1.0. However, the abundance gradients relative to the mean stellar composition are different for the four investigated elements. The largest gradient (1.3~dex) is found for Y. Sr shows less of a contrast (0.6~dex), which is comparable to that of Ti (0.5~dex). Finally, a very low contrast (0.15~dex) is observed for Cr.

There is a significant variation in the latitudinal distribution of chemical elements. In particular, the relative abundances are different at the visible rotational pole. For Cr we observe an underabundance at the pole, whereas for Ti, Sr, and Y the overabundance spots extend to high latitudes in a very similar way. The Cr underabundance zone occupies the whole polar region, while those of Ti, Sr, and Y occupy roughly half of it.

One can also note an interesting variation in the overall surface area covered by the spots of different elements. A major part of the stellar surface  shows a relative Ti overabundance. A somewhat smaller area is covered by Y spots, and even smaller occupied by Cr inhomogeneities. Sr is concentrated mainly in a single spot.

\onlfig{9}{
\begin{figure*}[!t]
  \centering
 {\resizebox{13cm}{!}{{\includegraphics{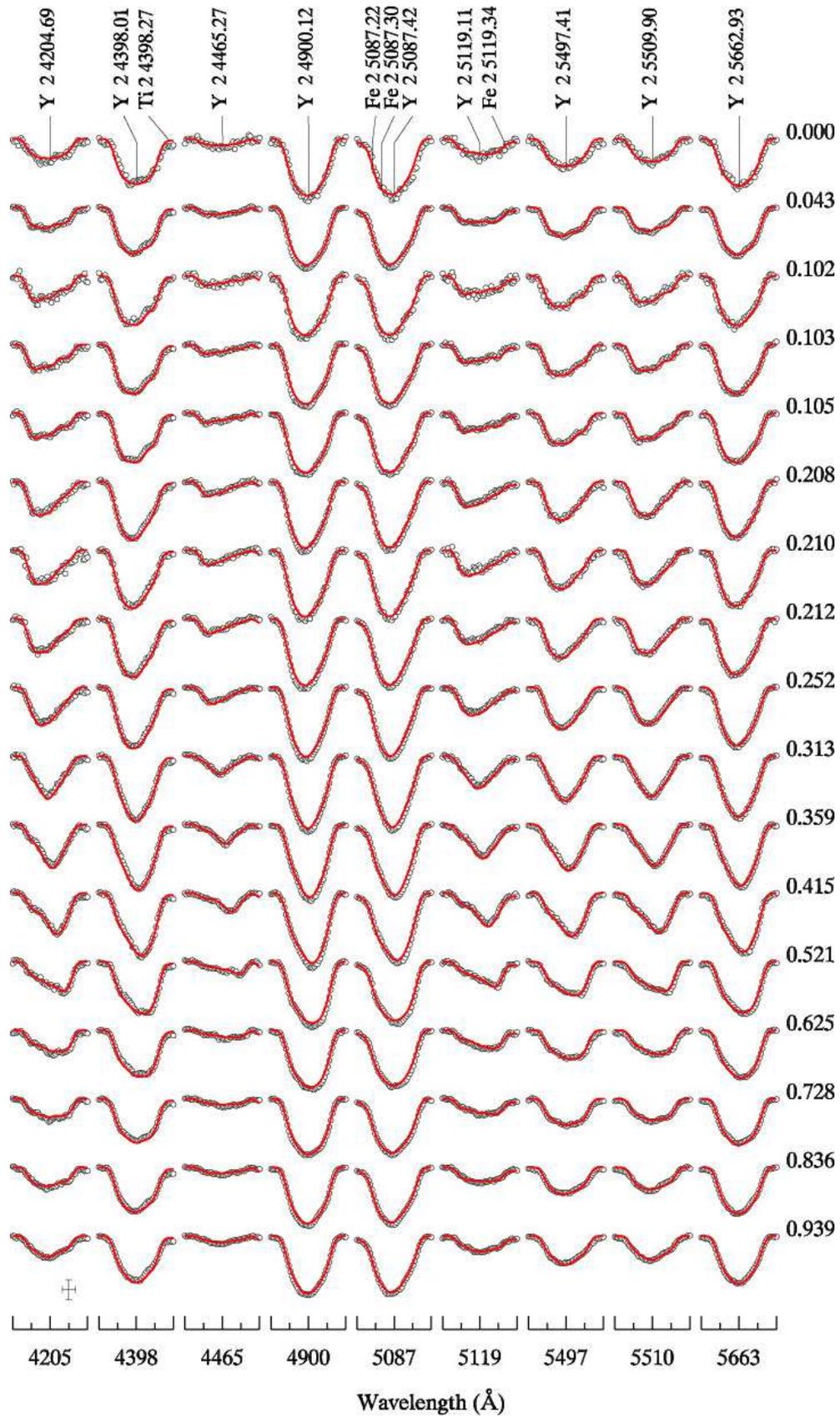}}}}
  \caption{Comparison of the observed ({\it symbols}) and model ({\it solid lines}) profiles for Y lines in \s. The bars in the lower left corner indicate the horizontal and vertical scales (0.1~\AA\ and 5\% of the continuum intensity, respectively).}
  \label{fitY}
\end{figure*}}

\onlfig{10}{
\begin{figure*}[!t]
  \centering
 {\resizebox{9cm}{!}{{\includegraphics[width=10cm]{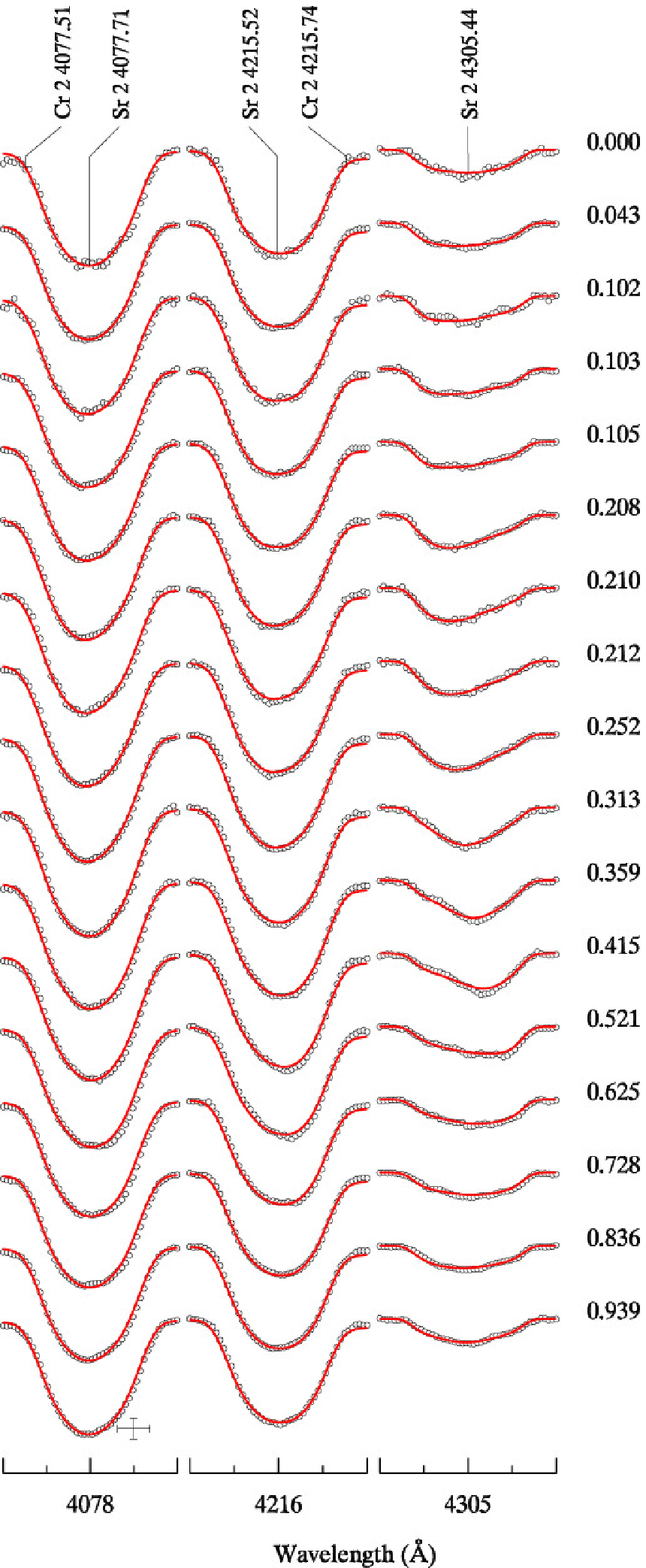}}}}
  \caption{Same as Fig.~\ref{fitY} but for Sr.}
  \label{fitSr}
\end{figure*}}

\onlfig{11}{
\begin{figure*}[!t]
  \centering
 {\resizebox{13cm}{!}{{\includegraphics[width=17cm]{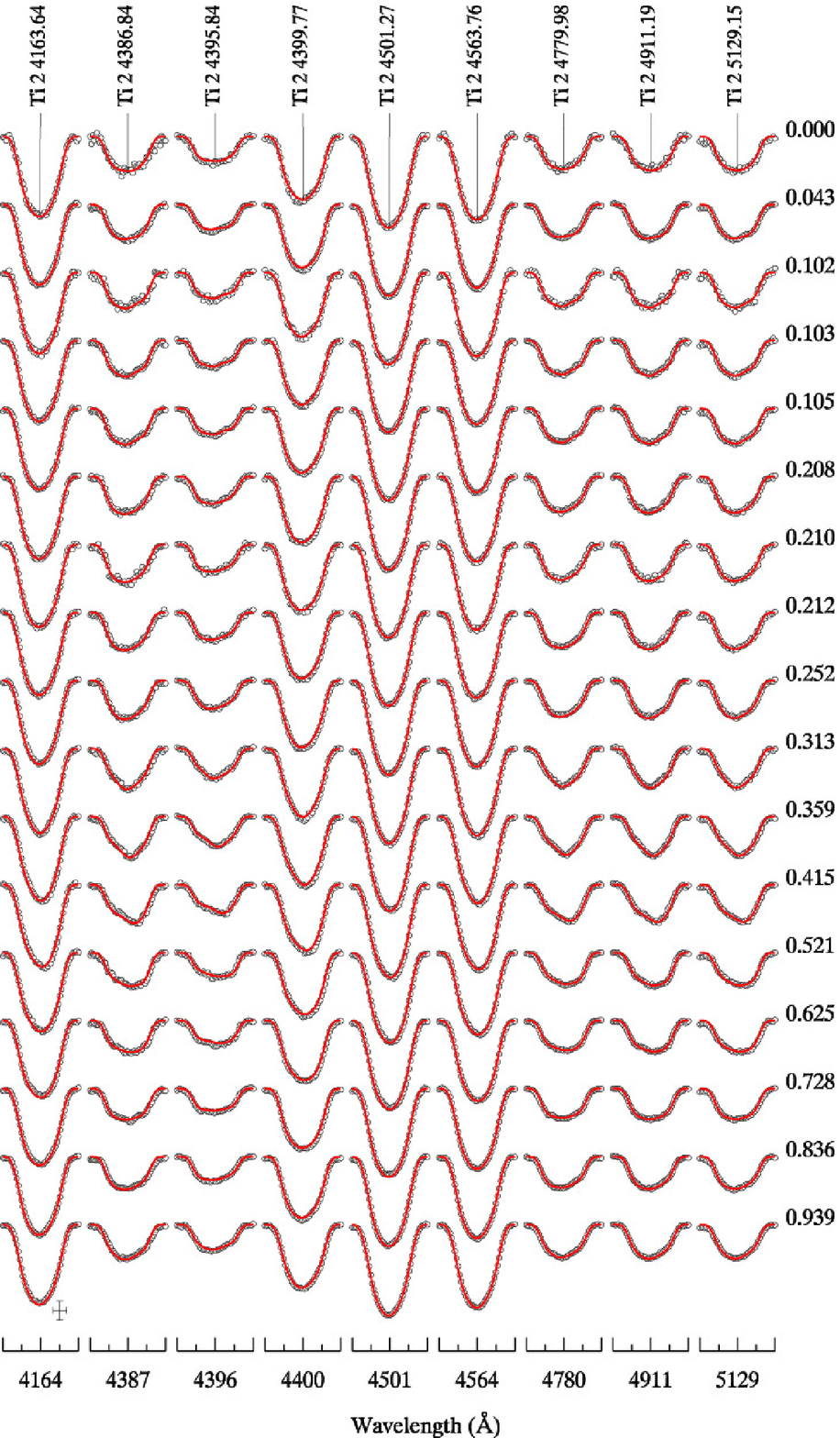}}}}
  \caption{Same as Fig.~\ref{fitY} but for Ti.}
  \label{fitTi}
\end{figure*}}

\onlfig{12}{
\begin{figure*}[!t]
  \centering
 {\resizebox{13cm}{!}{{\includegraphics[width=17cm]{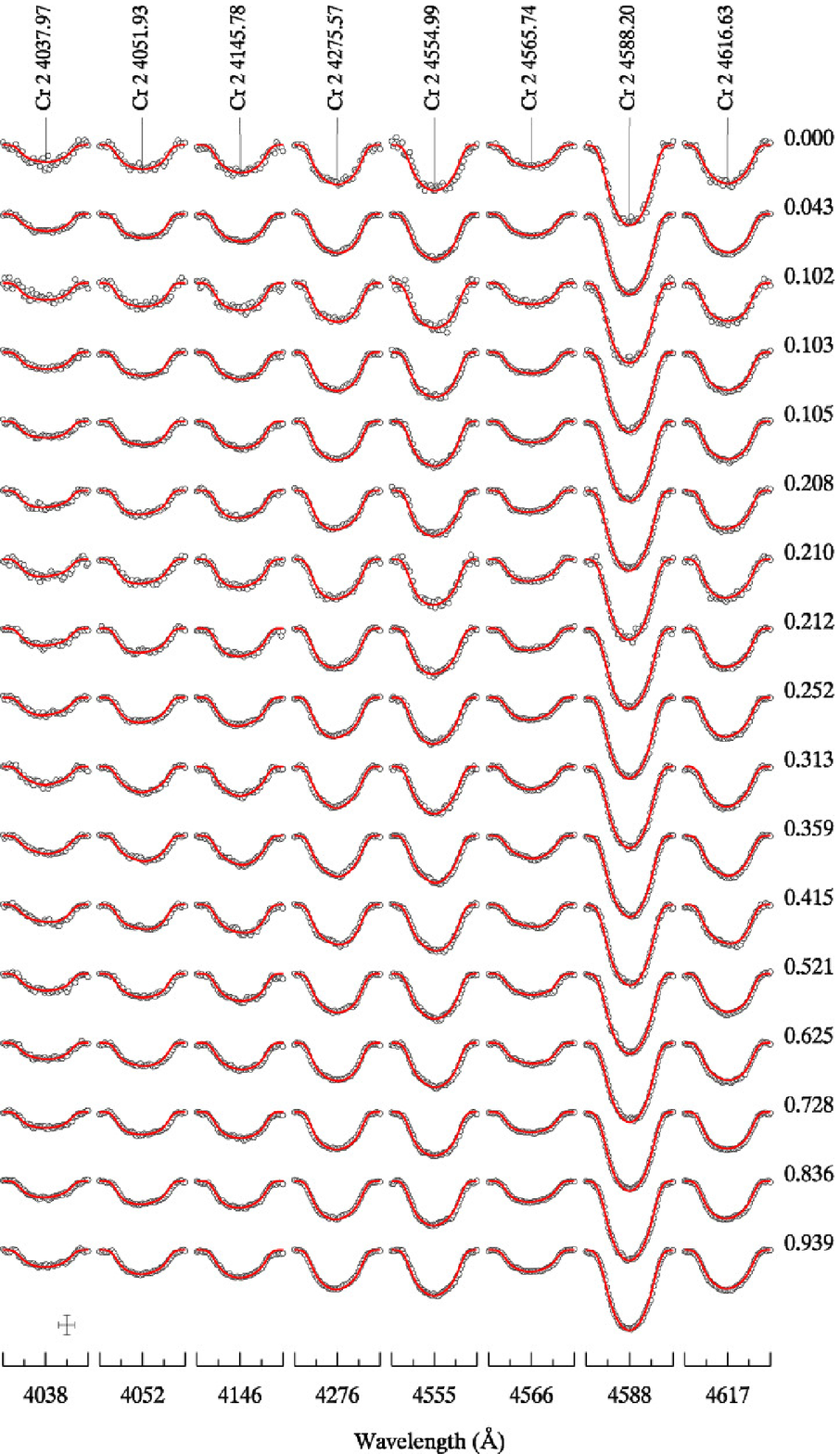}}}}
  \caption{Same as Fig.~\ref{fitY} but for Cr.}
  \label{fitCr}
\end{figure*}}

\section{Stratification of Y and Ti}
 \label{strat}
It is known that chemically peculiar stars may possess an inhomogeneous distribution (stratification) of some chemical elements with depth in stellar atmosphere \citep{Ryabchikova:2002}. In this case, computing a synthetic spectrum under the assumption of a homogeneous atmosphere will lead to  discrepancies between the observed and synthetic profiles of spectral lines formed at different atmospheric levels. However, such a discrepancy must be interpreted with caution due to a possible confusion with an inadequate quality of the adopted line transition data.

A complete stratification analysis of \s\ is beyond the scope of our paper. Instead, we aim at testing a very specific hypothesis that inhomogeneously distributed chemical elements are concentrated in the upper layers of the stellar atmosphere, where they might be affected by an extremely weak magnetic field. This scenario is inspired by the model of Hg stratification suggested by \citet{Michaud:1974} and by the recent discussion of the role of weak magnetic fields \citep{Alecian:2011}. We selected Y and Ti for the stratification analysis because their spots show a high relative abundance contrast, and these elements have a sufficient number of diverse spectral lines. The spectra for stratification analysis were obtained by averaging the line profiles utilised for DI.

We have carefully examined atomic data available for the selected \ion{Y}{ii} lines. Our primary source of transition probabilities is the revised \ion{Y}{ii} list kindly calculated by R.Kurucz\footnote{http://kurucz.harvard.edu/atoms/3901/}. These data are compatible with the oscillator strengths extracted from VALD, but include more lines. For seven out of nine \ion{Y}{ii} lines alternative $\log gf$ values are available from \citet{Biemont:2011}. We compared these two sets of oscillator strengths under the assumption of a homogeneous abundance, finding that the data from \citet{Biemont:2011} are less consistent internally. Since the lines giving discrepant results are not distinguished by their strength or excitation potential, this indicates a real limitation of the atomic data rather than an impact of possible stratification. The list by Biemont et al. also lacks information on two other lines, especially the high-excitation \ion{Y}{ii} 4465~\AA, which turned out to be very useful for stratification analysis.

On the other hand, all oscillator strengths of Ti lines were provided by VALD and came from a single source \citep{Pickering:2001}. The oscillator strength corrections provided in the last column of Table~\ref{DIlines} were not applied to Y and Ti lines in the stratification analysis. To compute stratification models, we used the IDL code DDAFit \citep{SYNTH3}, which provides an interface to the spectral synthesis program SYNTH3. DDAFit assumes a step-function parameterisation of the stratification profile. In this study we assumed a fixed sharp transition\,--\,0.2 on the $\log \tau_\mathrm{5000}$ scale\,--\,between the under- and overabundance zones and set the element abundance in the lower atmospheric to $\log(N_\mathrm{el}/N_\mathrm{tot})=-15$~dex. The abundance in the upper atmosphere was determined by the code by fitting observed profiles for a predefined set of abundance step positions between $\log \tau_\mathrm{5000}=-2$ to $-5$. Then the fit corresponding to each of the stratification models was compared to the one obtained with a homogeneous vertical distribution.

\begin{figure}[!t]
  \centering
 {\resizebox{\hsize}{!}{{\includegraphics{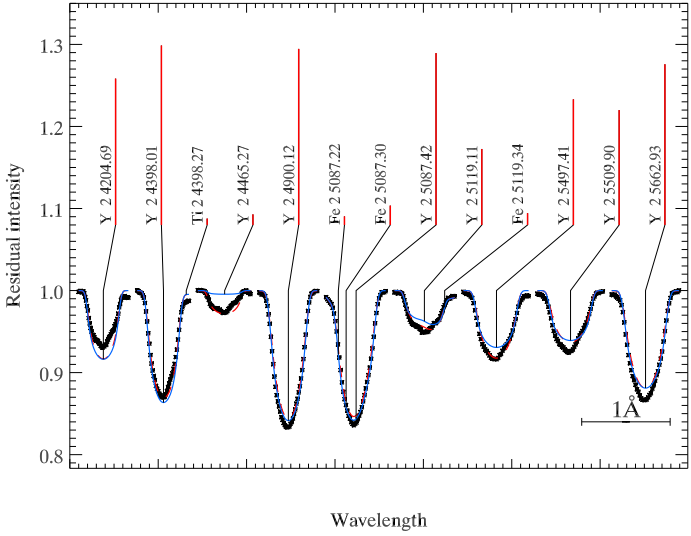}}}}
  \caption{Line profile fit for Y assuming a homogeneous abundance (dashed line) and stratification above $\log\tau_{5000}=-5$ (solid line). Observations are shown with asterisks.}
  \label{Fig10}
\end{figure}

In Fig.~\ref{Fig10} we illustrate the line profiles fits obtained for \ion{Y}{ii}. The fit provided by an extreme chemically stratified model yields worse agreement with the observations than does the model with a homogeneous Y distribution.

\begin{figure}[!t]
  \centering
 {\resizebox{\hsize}{!}{{\includegraphics{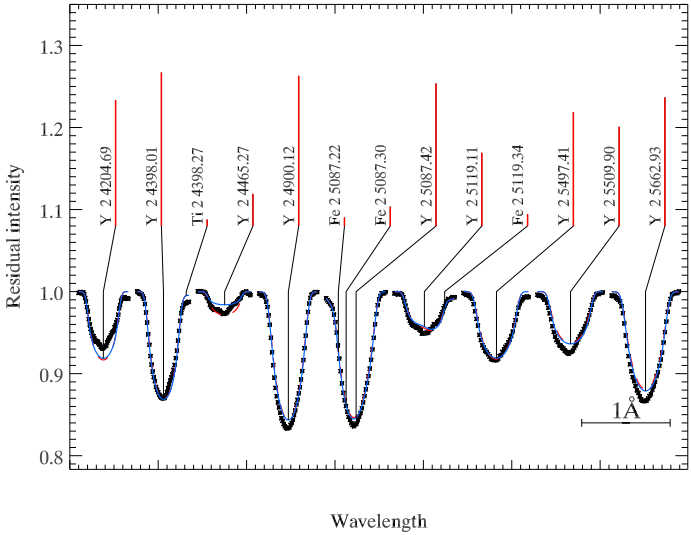}}}}
  \caption{Same as Fig.~\ref{Fig10}, but for the best-fit stratification model with a transition step at $\log \tau_\mathrm{5000}\approx-2$.}
  \label{Fig11}
\end{figure}

Standard deviation corresponding to different stratification profiles is reported in Table~\ref{tab6}. The first row in this table corresponds to a homogeneous distribution.  There is a marginal indication of Y stratification starting at $\log \tau_\mathrm{5000}\approx-2$. This is suggested by a slightly better fit of the observed line profiles, shown in Fig.~\ref{Fig11}.

\begin{figure}[!t]
  \centering
 {\resizebox{\hsize}{!}{{\includegraphics{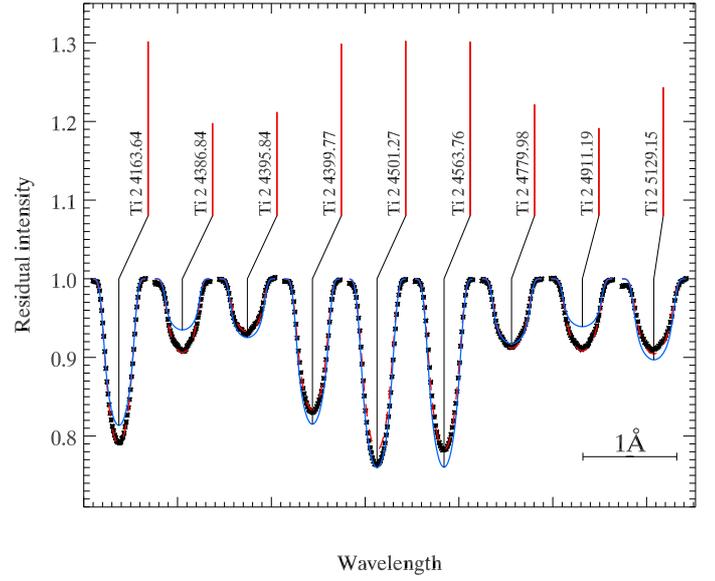}}}}
  \caption{Same as Fig.~\ref{Fig10}, but for Ti.}
  \label{Fig12}
\end{figure}

Ti line profiles were examined in the same way as Y. Figure~\ref{Fig12} illustrates the individual line profile fit in the format similar to Fig.~\ref{Fig10}. The assumption that all Ti is concentrated in the upper layers of the stellar atmosphere gives significantly worse results, compared to the homogeneous model. Similar to the case of Y, we found a marginal indication of Ti stratification with a step position at $\log \tau_\mathrm{5000}=-2$.

\begin{table}[t]
   \caption{Best-fit stratification models and corresponding standard deviations for Ti and Y.}
   \label{tab6}
   \centering
   \begin{tabular}{c|cccc}
  \hline\hline
         step at                 & \multicolumn{2}{c}{Ti}              & \multicolumn{2}{c}{Y} \\ \cline{2-5}
$\log \tau_\mathrm{5000}$ & $\varepsilon_\mathrm{upper}$ & $\sigma$ (\%)& $\varepsilon_\mathrm{upper}$ & $\sigma$ (\%)\\
  \hline
     & $-6.55$ & 0.536 & $-7.47$ & 0.693 \\
$-2$ & $-5.87$ & 0.523 & $-6.88$ & 0.648 \\
$-3$ & $-5.22$ & 0.680 & $-6.22$ & 0.667 \\
$-4$ & $-4.42$ & 0.941 & $-5.44$ & 0.757 \\
$-5$ & $-3.43$ & 1.192 & $-4.45$ & 0.874 \\
\hline
   \end{tabular}
   \tablefoot{The first row corresponds to a homogenous element distribution.}
\end{table}

\section{Summary and discussion}
 \label{disc}

Our study shows that \s\ does not possess a strong magnetic field. Owing to the quality of our spectropolarimetric data, we were able to set a sensitive upper limit of 4~G on the mean longitudinal magnetic field. Our results also do not support the hypothesis that magnetic field is associated with spots of chemical elements. From the spectral lines of inhomogeneously distributed chemical elements, we inferred the upper limit of 8\,--\,15~G for \Bz. As in the case of other HgMn stars studied with high-precision spectropolarimetry \citep{Auriere:2010, Makaganiuk:2011a}, the results of this work clearly demonstrate the absence of global magnetic field structures. Analysis of LSD profiles also rules out the presence of complex magnetic fields, similar to those found in active late-type stars \citep{Donati:2009}.

The full rotational phase coverage of our data enabled us to confirm the variability in the spectral lines of Y, Sr, and Ti, found earlier by \citet{Briquet:2010}. These authors also suggest a possible variability in Zr. We found a marginal indication of the variability in the \ion{Zr}{ii} line at 4149.2~\AA. However, an insufficient $S/N$ of our data in that spectral region, as well as a small number of visible Zr lines, does not allow us to conclusively investigate variability for this element. At the same time, a combination of superior data quality and the application of LSD technique enabled us to detect a very weak variability of Cr lines.

We applied the Doppler imaging technique to interpret the line profile variability of Y, Sr, Ti, and Cr. The main features of the inhomogeneous distribution of all these elements is a common spot located on one hemisphere of the star. This spot does not occupy the entire stellar hemisphere as in the case of another HgMn star 66~Eri~A studied by \citet{Makaganiuk:2011b}. Possibly, the morphological differences of chemical spot distributions observed in 66~Eri~A and \s\ are related to the absence of a close secondary component in the latter star.

Besides the large primary spots, we found two smaller ones for Ti and Cr. These chemical elements show a tendency to concentrate near the stellar equator, unlike Y and Sr, which show latitudinally extended spots. The spot morphology in HgMn stars is poorly studied due to the very small number of stars to which DI was applied so far. At the moment, there are only four HgMn stars, for which surface maps have been reconstructed ($\alpha$~And, AR~Aur~A, 66~Eri~A, and \s). By looking at their maps, one can note a tendency for chemical elements that are heavier than the mass number $Z\,\sim\,37$ to have more pronounced spots. These chemical elements also show a more contrast between the over- and underabundance zones.

We cannot directly confirm the evolution of chemical spots on \s\ based on comparing our maps with those published by \citet{Briquet:2010}. This is due to the differences in the data quality, the time span of observations, and the mapping approach used by Briquet et al. In particular, we used multiple spectral lines for Doppler imaging, which yields more robust surface distributions. Our surface maps are smoother and show less contrast, especially for Ti. In contrast, Briquet et al. finds a number of relatively small over- and underabundance spots of Ti. This may be related to insufficient quality of their spectra or incorrectly chosen regularisation parameter. Briquet et al. did not provide any individual line profile fits for Sr and Ti, which does not allow us to assess the reasons for the observed differences between our maps and theirs. However, we note that there appears to be a common overabundance zone at phases 0.8\,--\,0.9 in the Ti and Y maps obtained by \citet{Briquet:2010}. The common spot in our maps appears at phase 0.2. This difference may come from the differently assumed zero phase or insufficiently precise rotational period.

The evolution of chemical structures claimed by Briquet et al. generally consists of a redistribution of small-scale features. Our inversion results lack such local features entirely, suggesting that the spot evolution in \s\ may be an artefact of their poorly constrained Doppler reconstructions. The primary evidence for surface structure variation\,--\,the line profile and equivalent width difference at similar rotation phases\,--\,has to be examined for \s\ in the same way as for $\alpha$~And by \citet{Kochukhov:2007}.

It was suggested that heavy chemical elements, such as mercury, may accumulate in the uppermost atmospheric layers of HgMn stars \citep{Michaud:1974} where they could be affected by very weak magnetic fields \citep{Alecian:2011}. We tested this hypothesis for Ti and Y with the help of stratification analysis, assuming that these elements are concentrated only in the upper layers of the atmosphere of \s\ and are completely absent in the lower layers. This extreme stratification model clearly disagrees with  observations. Thus, we conclude that the horizontal element inhomogeneities in \s\ are not associated with high-lying chemical clouds. Previous stratification studies of HgMn stars \citep{Smith:1996, Thiam:2010} also found no evidence of any strong outward abundance increase. Nevertheless, in our analysis of \s\ the formally best fit is achieved for a stratification model with the transition between the over- and underabundance zones occurring at $\log \tau_\mathrm{5000}=-2$. This stratified distribution provides a marginally better description of the observed Ti and Y line profiles than a homogeneous abundance.

The absence of a strong magnetic field that could be associated with the formation of chemical spots suggests the operation of an alternative process of spot formation. That we observe evolving surface inhomogeneities in HgMn stars suggests a mechanism that affects the atomic diffusion, making it time-dependent. \citet{Alecian:1998} presents a qualitative scenario in which a chemical element is pushed by radiative acceleration to the upper layers of stellar atmosphere, forming an unstable stratification which changes with time. An increase in concentration of the element at the lower atmospheric layers absorbs photons, thus reducing the radiative acceleration received by the same element in the upper layers, leading to cyclic disappearance and accumulation of the upper overabundance zone.

In the recent paper, \citet{Alecian:2011} has performed a detailed numerical study of non-equilibrium, time-dependent diffusion for a fictitious chemical element similar to mercury. Their calculations suggest that the equilibrium stratification assumed by previous atomic diffusion studies \citep[e.g.,][]{LeBlanc:2009} may never be attained in reality and that vertical element stratification may become unstable whenever radiative acceleration is equal to gravity. Instabilities first develop in high layers and then propagate downwards. The diffusion time scale can be as short as one month at $\log\tau_{5000}=-4$, but increases to several years in deeper layers. The lack of concentration of Y and Ti in the highest layers combined with the claimed spot variation during two months seems to be too rapid compared to what could be expected from the numerical simulations by \citet{Alecian:2011}.

\begin{acknowledgements}
We thank T. Ryabchikova for the advice on atomic data and R. Kurucz for providing the new \ion{Y}{ii} line list. Comments by the referee, C. Folsom, have helped to improve the paper. OK is a Royal Swedish Academy of Sciences Research Fellow, supported by grants from Knut and Alice Wallenberg Foundation and Swedish Research Council.
\end{acknowledgements}

\bibliographystyle{aa}
\bibliography{references}

\end{document}